\documentclass[twocolumn,prb,showpacs,preprintnumbers,amsmath,amssymb]{revtex4-1}
\usepackage{dcolumn}
\usepackage{graphicx}
\usepackage{color}

\begin{document}

\author{V. M. L. Durga Prasad Goli}
\email{gvmldurgaprasad@sscu.iisc.ernet.in}
\affiliation{Solid State $\&$ Structural Chemistry Unit,
 Indian Institute of Science, Bangalore 560 012, India}

\author{Shaon Sahoo}
\email{shaon@physics.iisc.ernet.in}
\affiliation{Department of Physics,
Indian Institute of Science, Bangalore 560 012, India\\}

\author{S. Ramasesha}
\email{ramasesh@sscu.iisc.ernet.in}
\affiliation{Solid State $\&$ Structural Chemistry Unit,
Indian Institute of Science, Bangalore 560 012, India\\}

\author{Diptiman Sen}
\email{diptiman@cts.iisc.ernet.in}
\affiliation{Centre for High Energy Physics,
Indian Institute of Science, Bangalore 560 012, India}

\title{Quantum phases of dimerized and frustrated Heisenberg spin
chains with s = 1/2, 1 and 3/2: an entanglement entropy and fidelity study}

\begin{abstract}
{\noindent
We study here different regions in phase diagrams of the spin-1/2, spin-1 
and spin-3/2 one dimensional antiferromagnetic Heisenberg systems with 
frustration (next-nearest-neighbor interaction $J_2$) and dimerization 
($\delta$). In particular, we analyze the behaviors of the bipartite 
entanglement entropy and fidelity at the gapless to gapped phase transitions 
and across the lines separating different phases in the $J_2-\delta$ plane. 
All the calculations in this work are based on numerical 
exact diagonalizations of finite systems.}
\end{abstract}

\pacs{64.70.Tg, 75.10.Pq, 03.67.-a, 03.67.Mn}
\maketitle

\section{Introduction}

Matter can appear in different quantum phases with exotic properties like 
charge density wave, magnetism, superconductivity, and so on. Studies of 
these phases and the transitions from one phase to the other are 
important and interesting for both academic and technological 
reasons. Two major tools from quantum information theory have been used 
extensively in recent years for studying the quantum phases of a system: 
quantum entanglement and fidelity. The idea of quantum entanglement originated
in the study of quantum correlations of many-body systems 
\cite{horodecki,amico}. It is expected 
that even for moderately large system sizes, the entanglement entropy can 
identify the values of the parameters of the Hamiltonian where a quantum 
phase transition (QPT) occurs because the quantum correlations of the systems 
change significantly occurs when one goes across such a transition 
\cite{sachdev}. In recent years, entanglement entropy has been used to study 
quantum critical regions in various systems \cite{latorre,jin04,refael,
larsson,barthel06,laflo,calabrese09,tagliacozzo08,pollmann09,latorre09}. 
Quantum fidelity is a measure of how little the ground state of a system 
changes as one changes the parameters of the Hamiltonian. A large change 
in the fidelity is anticipated close to a QPT even if the system size is 
not very large
\cite{znidaric03,zanardi06,buon07,venuti07,giorda07,shigu08,ma08,
zhou081,schwandt09,eriksson09,gritsev09,polkovnikovrmp,sirker,rams11,mukherjee}.

In this paper, we study different quantum phases and quantum critical regions 
with frustration $J_2$ (next-nearest-neighbor coupling) and dimerization 
$\delta$ (an alternation in the nearest-neighbor couplings) of the spin-1/2, 
1 and 3/2 Heisenberg antiferromagnetic systems in one dimension by 
calculating the entanglement entropy and fidelity in the ground state of 
the system. In particular, we study the transition from a gapless to 
a gapped phase and the changes in the spin structure across different 
phase lines in the $J_2-\delta$ plane for the spin-1/2, 1 and 3/2 systems. For
the spin-1/2 and spin-1 systems, our numerical study contains the locations of
different critical points and lines separating different phases in the phase 
diagrams which had been found earlier by other methods, such as the density 
matrix renormalization group method 
\cite{white1,chitra,totsuka,pati1,pati2,white2}. 
We compare our results with those reported previously (using different 
techniques) whenever possible. For the spin-3/2 system, we find that the 
entanglement entropy and fidelity helps us to estimate the locations of 
the various critical points and lines in the phase diagram.

\begin{figure*}[]
\begin{center} {\includegraphics[width=16.0cm]{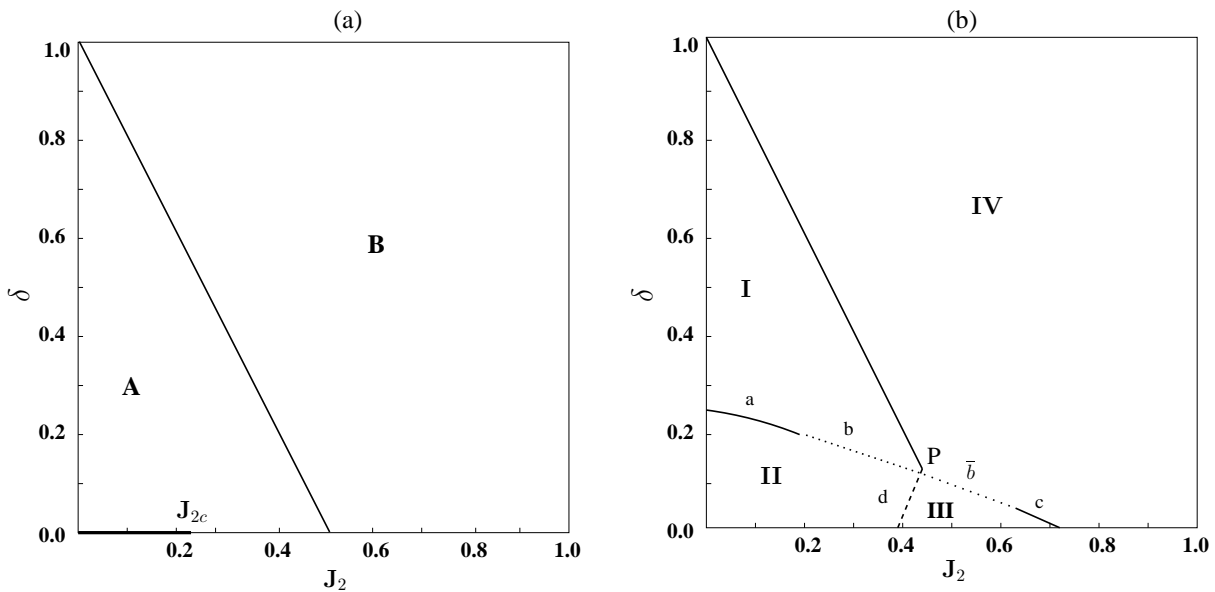}}
\caption{\small Phase diagrams of the (a) spin-1/2 and (b) spin-1 chains in 
the $J_2-\delta$ plane.} \label{phase_diag} \end{center} \end{figure*}

There are some related works which we briefly mention here. The bipartite 
entanglement for the spin-1/2 $J_1-J_2$ model (without dimerization) has been 
studied in different contexts, like the transition from the Neel to the spiral
phase and the gapless to gapped phase transition along the $J_2$-axis (see, 
for example, \cite{alet,ravi}). The gapless to gapped phase transition has 
also been studied using the fidelity (as a function of $J_2$) of the first 
excited state with periodic boundary conditions \cite{chen} and
the fidelity susceptibility of the ground state \cite{thesberg}. The role of 
entanglement between distant sites has been studied for this model with 
frustration and dimerization \cite{campos}. To the best of our knowledge, the 
phase diagram of the spin-1 and spin-3/2 $J_1-J_2$ model (with or without 
dimerization) has not yet been studied using entanglement entropy and fidelity. 
In this paper we report results on the entanglement entropy and fidelity of 
the $J_1-J_2-\delta$ model over the $J_2-\delta$ plane.

Our paper is organized in the following way. In section II, we discuss how to 
calculate the entanglement entropy and fidelity. We then introduce the 
Hamiltonian that we study in section III. In section IV, we give a 
brief introduction to the numerical techniques we use in our work. We then 
present the entropy and fidelity results for the spin-1/2, spin-1 and spin-3/2
systems in section V. We conclude our paper in section VI.

\section{Entanglement Entropy and Fidelity}

A pure state of a bipartite entangled system can be written as 
$\displaystyle|\psi\rangle = \sum_{ij} C_{ij} |\phi_i\rangle^l |\phi_j
\rangle^r$, where $ |\phi_i\rangle^l$ and $|\phi_j\rangle^r$ are the basis 
states of the left and right blocks respectively. The reduced density matrix 
(RDM) of the left block, $\rho_l=Tr_r(|\psi\rangle \langle\psi|)$, is 
calculated by tracing out the degrees of freedom of the right block.
The elements of the RDM $\rho_l$ are given by 
\begin{eqnarray} \rho_{ij} = \sum_k C_{ik}C_{jk}^* \label{rho1}. 
\end{eqnarray}
The von Neumann entropy of a block is given by $ S=-Tr(\rho~log_2~\rho )$ or 
\begin{eqnarray} S = -\sum_i ~\lambda_i~{\rm log_2}~\lambda_i, \label{entrp2} 
\end{eqnarray}
where the $\lambda_i$'s are the eigenvalues of $ \rho$.

Fidelity measures how little a particular wave function (for instance, the 
ground state) changes with the parameters of a model Hamiltonian. This is 
quantified by the overlap of the wave function at two different parameter 
values. If $p$ is a parameter then the fidelity is given by
\begin{eqnarray} 
F=|\langle\Psi(p)|\Psi(p+\alpha)\rangle|, \label{fidelity1}
\end{eqnarray}
where $\alpha$ is a small variation in $p$. In our case, both $J_2$ and 
$\delta$ are parameters with respect to which we have calculated fidelity 
and we have taken the change in the parameter to be $10^{-2}$ in the 
numerical calculations.

\section{Description of the Spin Model}

We study the Heisenberg Hamiltonian for the antiferromagnetic chain with 
both nearest-neighbor and next-nearest-neighbor couplings and dimerization 
\cite{majumdar,chitra}. We will use this Hamiltonian for the spin-1/2,
spin-1 and spin-3/2 systems; it is given by
\begin{eqnarray} 
H={J_1} \sum_{i=1}^{2N-1} (1-{(-1)^i}~\delta) \vec{S}_i\cdot \vec{S}_{i+1}+
\displaystyle {J_2}\sum_{i=1}^{2N-2} { \vec{S}_i\cdot \vec{S}_{i+2}}, 
\label{hspin} \end{eqnarray}
where $J_1$ is the nearest-neighbor interaction (we take $J_1$=1 for our 
study), $\delta$ ($0\leq \delta \leq 1$) is the dimerization and $J_2$ 
($0\leq J_2 \leq 2$) is the next-nearest-neighbor interaction. In our 
entropy contour plot we have taken the range of $J_2$ from 0 to 2.

The phase diagrams of the spin-1/2 and spin-1 chains are different from 
each other in the $J_2-\delta$ plane (see Fig. \ref{phase_diag}). The spin-1/2 
system undergoes a QPT from a gapless phase to a gapped phase at $J_{2c}=
0.2411\pm0.0001$ without dimerization ($\delta=0$) \cite{okamoto}, while the 
rest of the phase diagram is gapped. The line $2J_2+\delta=1$ separates the 
Neel phase (region A) from the spiral phase (region B) \cite{chitra}.

\begin{figure}[]
\begin{center} {\includegraphics[width=9.0cm]{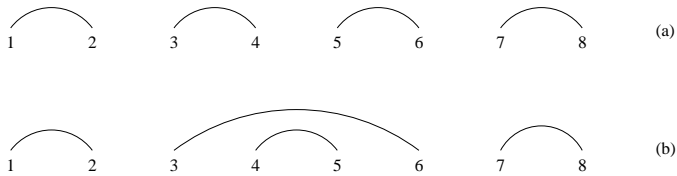}}
\caption{\small Some legal valence bond diagrams of a spin-1/2 chain with 
8 sites and total spin S = 0.} \label{sites1} \end{center} \end{figure} 

The spin-1 system has a number of distinct phases in the $J_2-\delta$
plane. Region I denotes the spin Peierls gapped phase. Regions II and III 
are Haldane gapped and spiral regions respectively. In these two regions 
the ground state of an open chain is four-fold degenerate. Region 
IV is a spiral gapped phase with a non-degenerate ground state for an open 
chain. A gapless phase exists along the critical line `a' that lies between 
(0, 0.25) to (0.22$\pm$0.02, 0.20$\pm$0.02) in the $J_2-\delta$ plane 
\cite{pati1,pati2,sierra}. A line `c' separating the regions II and III 
extends from (0.73, 0) to (0.65, 0.05); on that line the gap appears to be 
zero (to numerical accuracy). Along the dotted lines `b' which extends from 
(0.22$\pm$0.02, 0.20$\pm$0.02) to the point P = (0.432, 0.136) and 
`$\overline{b}$' which extends from P to (0.65, 0.05), the gap shows a 
minimum as a function of $\delta$ (see Fig. \ref{spin1_energy_lineb}). 
This will be discussed more detail in section V.B.2. The line $2J_2+\delta=1$ 
starts at the point `P' and extends up to (0, 1), separating regions I and IV. 
Another line (`d') starts at (0.39, 0) and ends at point `P', separating 
regions II and III.

The phase diagram of spin-3/2 system has not been studied yet. We will 
show below that the entanglement entropy and fidelity of this system can 
give some insights into its phase diagram.

\section{Numerical Techniques}

For our calculations, we use the $M_s$ basis (eigenstates of the $z$ 
component of total spin) \cite{pati3}. These basis states are orthonormal 
and it is easy to obtain the RDM (which is used to calculate the 
entanglement entropy) when the states of the system are expressed 
in this basis. On the other hand, most of the results can be understood
qualitatively using valence bond (VB) theory \cite{prasad}. In this theory 
\cite{pauling,soos}, the basis states in the singlet space are expressed as 
products of pairwise singlets, which follow the Rumer-Pauling rules, to 
avoid overcompleteness of the VB states. 
Fig. \ref{sites1} shows some VB-diagrams of a spin-1/2 chain with 8 sites 
and total spin $S = 0$. The VB state (a) is a Kekule state which is a product 
of nearest-neighbor singlets.

\begin{figure}[]
\begin{center} {\includegraphics[width=8.5cm]{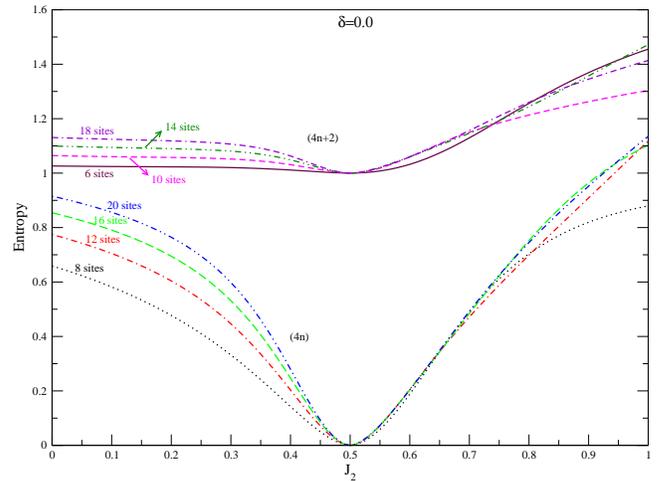}}
\caption{\small Ground state entanglement entropy of the spin-1/2 system 
with different chain lengths at $\delta=0$. The lower set of plots for 4n 
systems and the upper set of plots is for 4n+2 systems.} \label{entr_j2del00}
\end{center} \end{figure}

\begin{figure}[b]
\begin{center}
{\includegraphics[width=8.2cm]{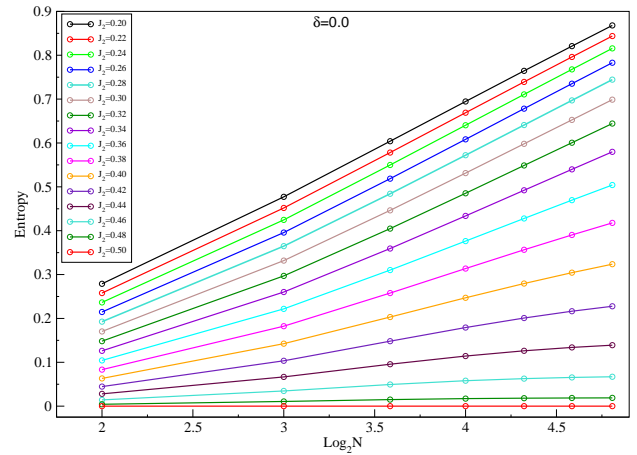}}
\caption{\small For a spin-1/2 system, the entropy versus 
logarithm of the system size is shown for different $J_2$ values for a 
uniform chain. 
\label{scale_spin1by2}} \end{center} \end{figure}

\begin{figure*}[]
\begin{center} {\includegraphics[width=16.0cm]{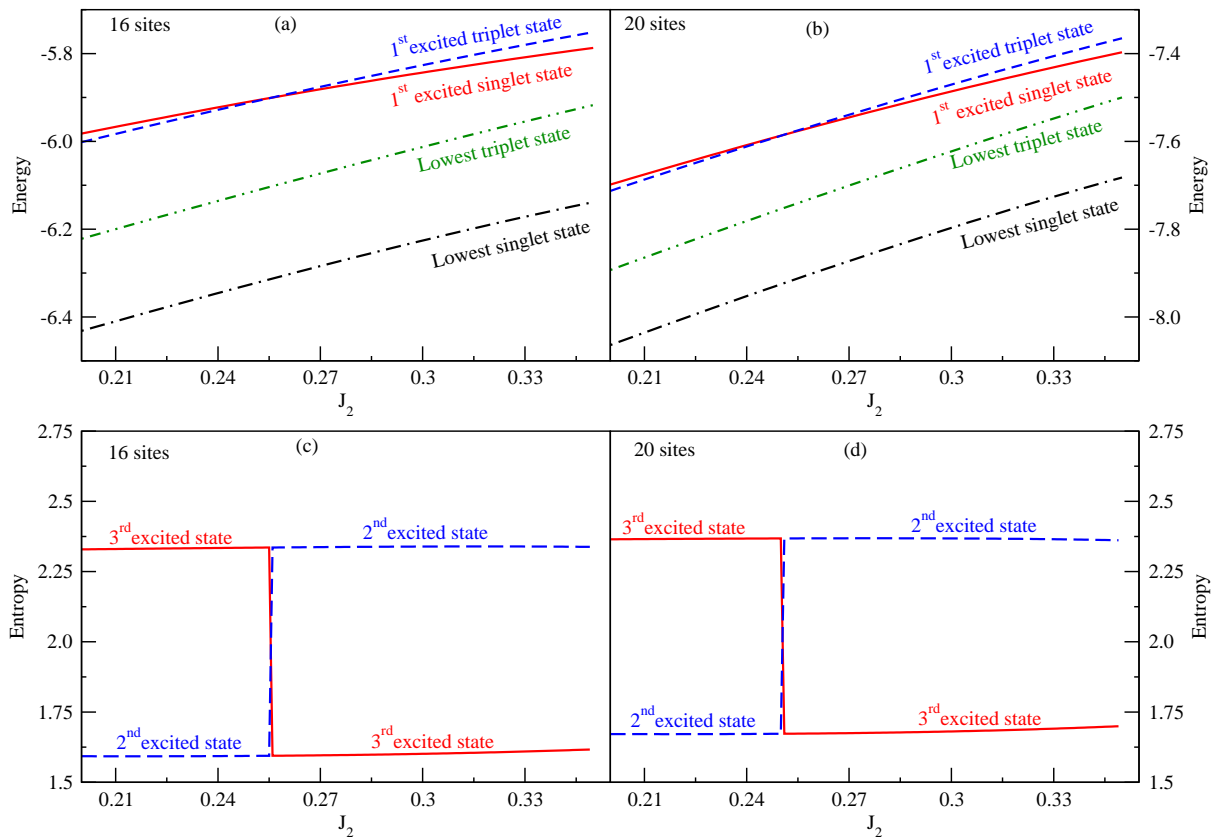}}
\caption{\small Lowest and first excited state energies of the 16 and 20 
sites spin-1/2 chain for different $J_2$ values with total spin S=0 and 1 
are shown in (a) and (b). Entropy of the second excited state (dashed line) 
and third excited state (solid line) of 16 and 20 sites for different $J_2$ 
values in the $M_s=0$ sector are shown in (c) and (d).} \label{J2crossing1} 
\end{center} \end{figure*}

Since our Hamiltonian conserves total spin, all its eigenstates are also 
eigenstates of total spin. Therefore, the eigenstates that we obtain by 
diagonalizing the Hamiltonian in the constant $M_s$ basis will be 
linear combinations of the VB basis states. It may be worth mentioning here
that appropriate linear combinations of constant $M_s$ basis states can 
give different VB basis states \cite{sahoo}. Now, a VB basis state 
contributes to the bipartite entanglement entropy of a state under study if 
the boundary between the two blocks of the system cuts a singlet line. If the 
boundary does not cut a singlet line, its contribution is assumed to be zero. 
For example, if the boundary goes through the sites 4 and 5 of the system, the 
entropy contribution of diagram (a) will be zero while that of diagram (b) 
will be be non-zero (Fig. \ref{sites1}). Depending upon the entropy 
contributions of the VB basis states and their relative weights in a state 
under study, we can qualitatively understood the entropy of the state 
\cite{prasad}. In generating the contour plot of entropy, we have calculated 
the entropy over a grid of 201 $J_2$ values (0$\le$$J_2$$\le$2) and 101 
$\delta$ values (0$\le\delta\le$1).

\section{Results and Discussion}

In this section, we present numerical results for the entanglement entropy of 
finite size chains with two equal block sizes for the spin-1/2, 1 and 3/2.
We also present results for the ground state fidelity of both the systems in 
the $J_2-\delta$ plane. 

\subsection{The spin-1/2 system in $J_2-\delta$ plane}

\subsubsection{Uniform chain ($\delta=0$)}

The uniform spin-1/2 chain without dimerization ($\delta=0$) goes through a 
gapless to gapped phase transition at $J_{2c} \simeq$ 0.2411 in the 
thermodynamic limit ($N\to\infty$) \cite{okamoto}, and its 
Neel phase is separated from the spiral phase at $J_2=0.5$. To study the 
behavior of the system around those points, we calculate the ground state 
fidelity and entanglement entropy of the system.

\begin{figure}[b]
\begin{center} {\includegraphics[width=8.5cm]{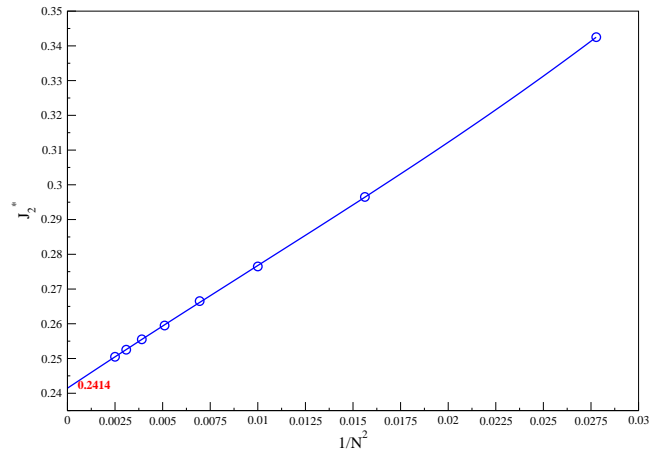}}
\caption{\small Convergence of the crossing points of the excited states for
a spin-1/2 chain.} \label{convergence_j2} \end{center} \end{figure}

\begin{figure*}[]
\begin{center} {\includegraphics[width=14.0cm]{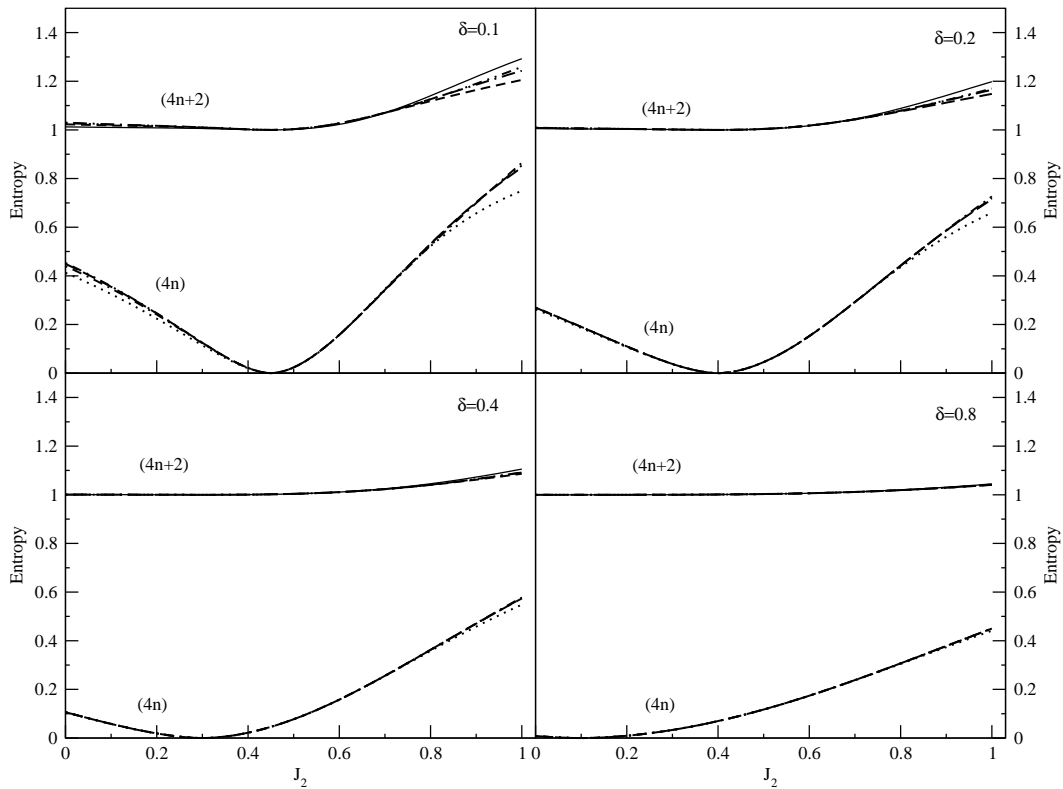}}
\caption{\small Ground state entanglement entropy of a spin-1/2 chain with 
lengths of 6, 8, 10, 12, 14, 16, 18 and 20, for different values of $\delta$ 
values; the line types for different lengths are the same as in Fig. 
\ref{entr_j2del00}. The upper set of curves is for chain length $4n+2$ while 
the lower one is for chain length $4n$ ($n$ being a positive integer).
Near the minimum, the finite size dependence is very weak.} 
\label{entropy_j2del} \end{center} \end{figure*}

We first consider the ground state entanglement entropy of finite size 
spin-1/2 chains with different $J_2$ values. The bipartite entropy for 
different chain lengths (and equal block size) can be seen in Fig. 
\ref{entr_j2del00}. At $J_2=0.5$, the entropy reaches a minimum; away from 
that point the entropy increases. For the systems with even block sizes, the 
minimum of the entropy goes to zero while for odd block sizes, this minimum 
is one. This result can be explained by noting that, at $J_2=0.5$ (the 
Majumdar-Ghosh point \cite{majumdar}), the ground state has Kekule state 
structure (as in Fig. \ref{sites1} (a)). Depending upon the block size being 
odd or even, entropy will be finite or zero respectively \cite{prasad}. As 
$J_2$ moves away from that point, the presence of other VB basis states 
(as in Fig. \ref{sites1}(b)) in the ground state will become significant. 
Since these VB basis states will have a finite entropy contribution, the 
entropy of the ground state will increase as $J_2$ moves away from the point.

\begin{figure*}[]
\begin{center} {\includegraphics[width=16.0cm]{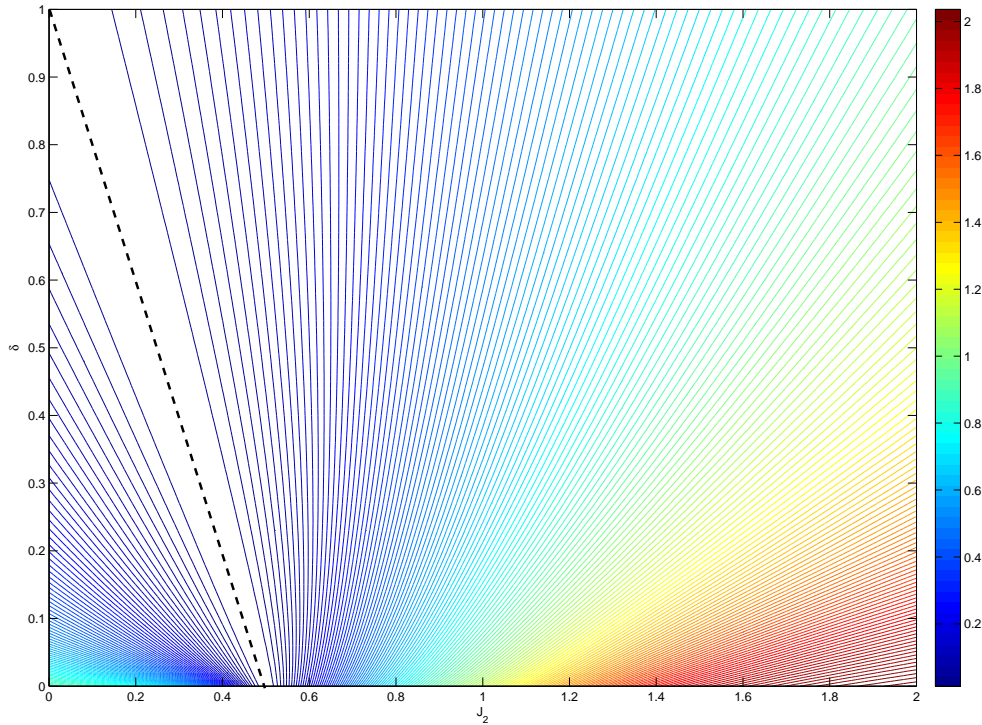}}
\caption{\small Ground state entanglement entropy contour plot of spin-1/2 
chain with 20 sites, for $J_2$ going from 0 to 2 and $\delta$ going from 
0 to 1.} \label{spin1by2_contour1} \end{center} \end{figure*}

We do not observe any change in the behavior of the ground state 
entropy around $J_{2c}$. To investigate this further, we plot the entropy 
versus $\log_2 N$ for different $J_2$ values (see Fig. \ref{scale_spin1by2}). 
The plots indicate that the present system sizes are too small to numerically 
verify the conformal field theory prediction of $S=\frac{c}{6}\log_2 N$ 
(with $c=1$) at $J_{2c}$ \cite{calabrese09}.
The first excited states in the singlet and triplet sectors cross 
as a function of $J_2$. We have calculated the entropy of these two states 
as a function of $J_2$. We find that the entropy of a state depends only 
on its spin and not on its energy. Therefore as a function of 
energy level ordering the entropy shows a jump at a value of $J_2^*$ 
which depends on the chain length. The jump in the value of the entropy can 
be seen from Fig. \ref{J2crossing1}. 
In Figs. \ref{J2crossing1} (a) and (b), we see that the ground state (lowest 
singlet) and lowest triplet state are non-degenerate for finite N; they become 
degenerate in the thermodynamic limit \cite{chitra,okamoto}. However, the first 
excited singlet and the first excited triplet states become degenerate near 
$J_{2c}$ even for small values of N. In Fig. \ref{convergence_j2} we plot the 
$J_2^*$(N) as a function of 1/$N^2$ and we see that $J_2^*$ extrapolates to 
$J_{2c}$ in the thermodynamic limit. The extrapolated value of $J_2^*$ 
(0.2414) is very close to the reported value of $J_{2c}=0.2411$
\cite{chitra,okamoto}. 

Upon calculating the ground state fidelity (taking $J_2$ to be the variable 
parameter), we do not observe any behavioral change at $J_2$ = 0.5 or 
$J_{2c}$. The fidelity of first excited states in both the singlet and the 
triplet sectors falls to zero near $J_{2c}$. The same thing has already 
been observed for the system with periodic boundary conditions \cite{chen}.

\subsubsection{Dimerized chain ($0<\delta\leq1$)}

In the phase diagram of the spin-1/2 system with dimerization, the Neel phase 
is separated from the spiral phase by the line $2J_2+\delta=1$. We study the 
ground state entanglement entropy of finite size chains in these phases 
with different $\delta$ values (see Fig. \ref{entropy_j2del}). We 
observe that the entropy of the system is minimum for values 
of $J_2$ and $\delta$ which fall on this line. For systems with even sized 
blocks, this minimum value is zero while for odd sized blocks the minimum is 
one. The reason for this is similar to the case $\delta = 0$ as given earlier. 

We have calculated the ground state fidelity (with $J_2$ as the variable
parameter) of finite size systems in the $J_2-\delta$ plane;
 we do not observe any sudden change in fidelity along the $2J_2+\delta=1$ 
line. This can be explained by the fact that the phases of the system on 
both sides of the line are gapped which implies that the ground state 
does not cross any excited state (i.e., the ground state does not change 
its character of being a singlet or a triplet) when we cross this line in 
the parameter space. 

\subsubsection{Spin-1/2 entropy phase diagram (contour plot)}

\begin{figure*}[]
\begin{center} {\includegraphics[width=13.0cm]{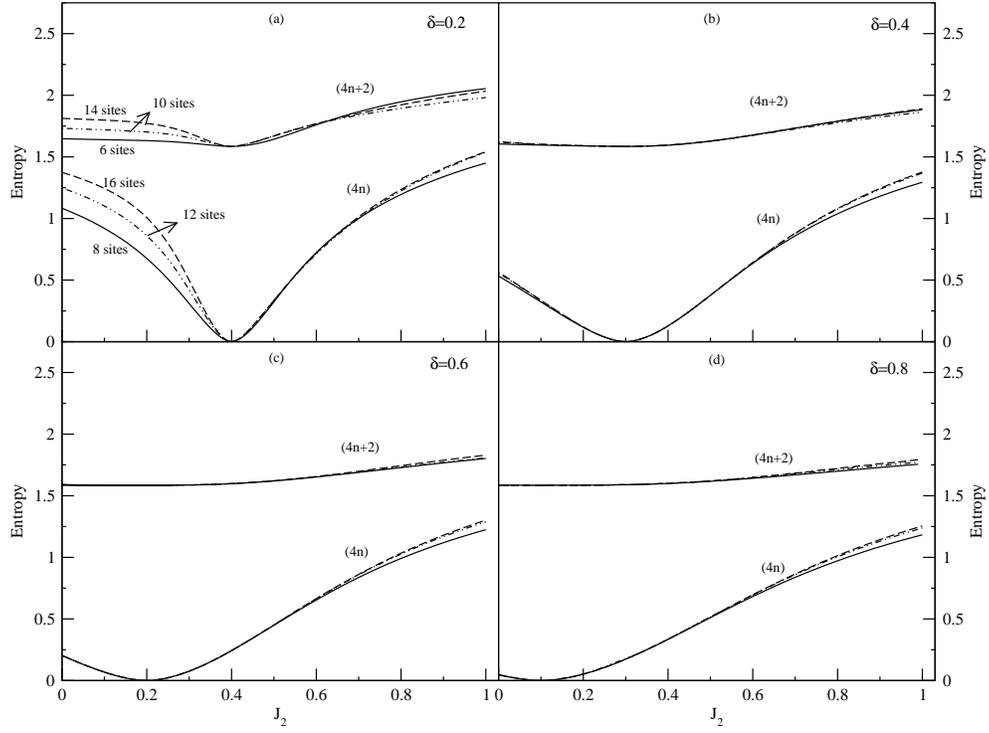}}
\caption{\small Ground state entropy of the spin-1 system with different chain
lengths and $\delta$ values along the $J_2$ axis. The line types for different 
chain lengths are the same as in (a).} \label{spin1_ent2} 
\end{center} \end{figure*}

\begin{figure*}[]
\begin{center} {\includegraphics[width=14.0cm]{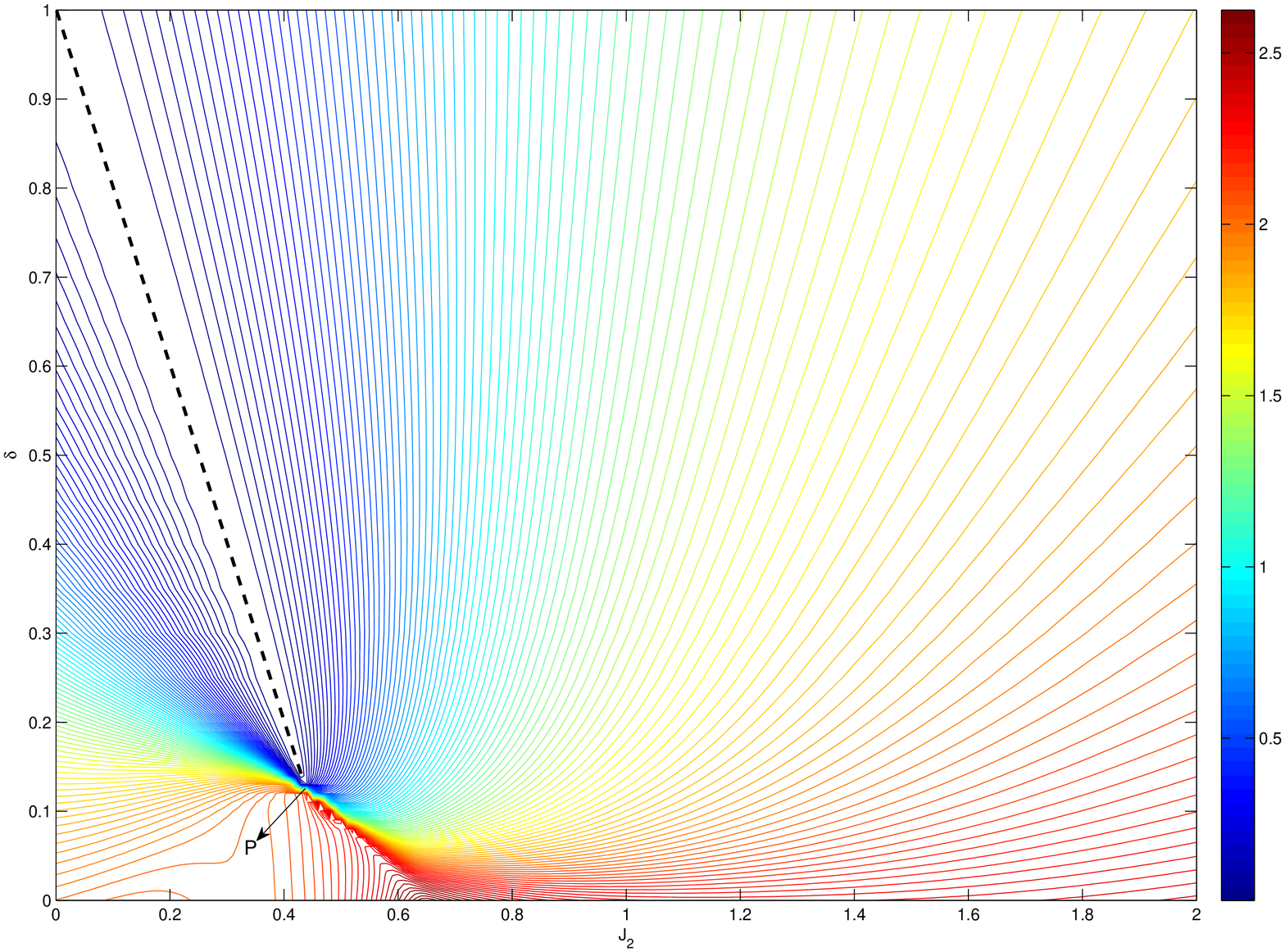}}
\caption{\small Ground state entanglement entropy contour plot of a spin-1 
chain with 16 sites, with $J_2$ going from 0 to 2 and $\delta$ going from
0 to 1.} \label{spin1_contour1} \end{center} \end{figure*}

\begin{figure}[b]
\begin{center} {\includegraphics[width=8.5cm]{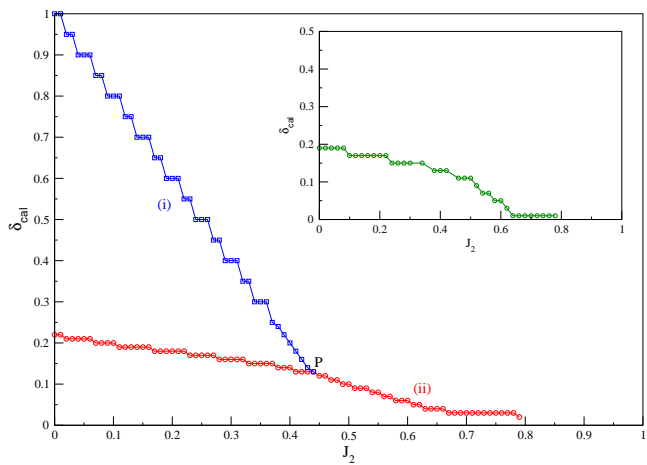}}
\caption{\small Curve (i) shows the $(J_2,\delta_{cal})$ values (squares) 
corresponding to the minimum entropy. Curve (ii) shows the 
$(J_2,\delta_{cal})$ values (circles) corresponding to the maximum absolute 
values of the first order derivative of the entropy w.r.t. $\delta$. 
Inset shows the points corresponding to minimum fidelity (taking $\delta$ 
as the variable parameter). All the results are obtained for the 16 site 
spin-1 chain.} \label{spin1_derv1} \end{center} \end{figure}
 
\begin{figure*}[]
\begin{center}
{\includegraphics[width=14.5cm]{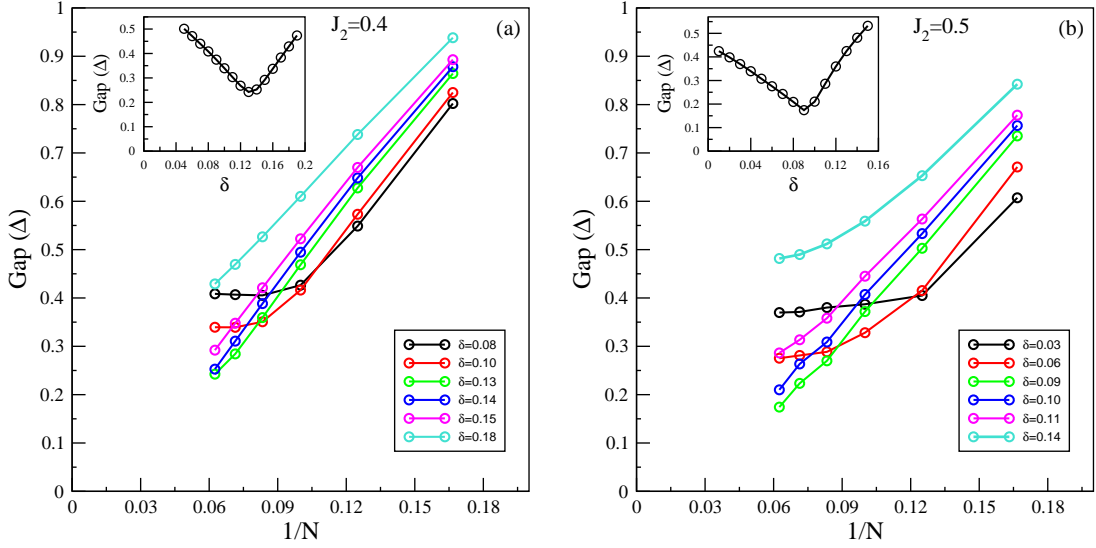}}
\caption{ \small 
Figures show how a plot of the gap ($\Delta$) versus $1/N$ changes with 
$\delta$ for (a) $J_2$=0.4 and (b) $J_2$=0.5 which lie on the dotted lines 
`b' and `$\overline{b}$' respectively in Fig. \ref{phase_diag}(b).
Here the gap represents $\Delta$ = $E_1$($M_s=1$)-$E_0$($M_s=0$). 
Insets show how the gap varies with $\delta$ for the 16 site spin-1 chain. 
See section V.B.2 for details.}
\label{spin1_energy_lineb} \end{center} \end{figure*}
 
We study the gapless to gapped phase transition and change in the spin 
structure (order-disorder change) along the $2J_2+\delta=1$ line using the 
entropy contour plot for a spin-1/2 chain with 20 sites (see Fig. 
\ref{spin1by2_contour1}). The entropy is zero along this line. 
The density of the entropy contour lines shows whether the system is in a
gapless or gapped phase. From this figure we see that the gapless 
region between $J_2=0$ and $J_{2c}$ has a higher density of entropy contour 
lines. In the rest of the figure the density of the contour lines 
is lower which shows that the rest of the phase diagram is gapped. At higher 
values of $J_2$, the spin-1/2 chain behaves like two decoupled chains with a 
weak coupling ($J_1$) between them. The stronger interaction ($J_2$) is 
responsible for the higher entropy. For higher values of both $J_2$ and 
$\delta$ values, the spin-1/2 chain behaves like a spin ladder with a 
gapped phase.

\subsection{The spin-1 system in $J_2-\delta$ plane}

As mentioned in section III, there are many phases in the $J_2-\delta$ phase 
diagram of the spin-1 system. To study these phases, we first calculate the
bipartite entanglement entropy in the ground state for finite systems.
We calculate the ground state entropy for different values of $\delta$ (with 
large $\delta$) and for different chain lengths (see Fig. \ref{spin1_ent2}). 
The entropy is zero along the line $2J_2+\delta=1$ for the systems with even 
sized blocks and non-zero for systems with odd sized blocks just as in the 
spin-1/2 case. As in the spin-1/2 case, the finite size effects are very weak 
near the minima of the entropy (Fig. \ref{spin1_ent2}).

\subsubsection{Spin-1 entropy phase diagram (contour plot)}

In Fig. \ref{phase_diag} (b), we know that the ground state is 
four-fold degenerate in regions II and III for an open chain. Hence we used 
the spin parity symmetry to break the degeneracy between 
the states corresponding to the total spin S = 0 and 1. Then we calculate the
entanglement entropy for the lowest state in the even parity subspace (which 
is a singlet). We study the quantum phases and QPTs of the spin-1 chain with 
16 sites using the contour plot of the ground state entropy 
in the $J_2-\delta$ plane, where $J_2$ goes from 0 to 2 and 
$\delta$ goes from 0 to 1 (see Fig. \ref{spin1_contour1}). 
We see in the figure that the line $2J_2+\delta=1$ starts approximately 
at the point P = (0.432, 0.136) and extends up to (0, 1). Along this line the 
entropy is zero. Along the gapless lines `a' and `c' (in Fig. 
\ref{phase_diag} (b)) the density of the entropy contour lines is higher, 
while the density is lower in the rest of the phase diagram. About the line 
`d' (in Fig. \ref{phase_diag} (b)), we observe that the density of contour 
lines is much lower compared to the regions near by. 

\begin{figure*}[]
\begin{center} {\includegraphics[width=15.0cm]{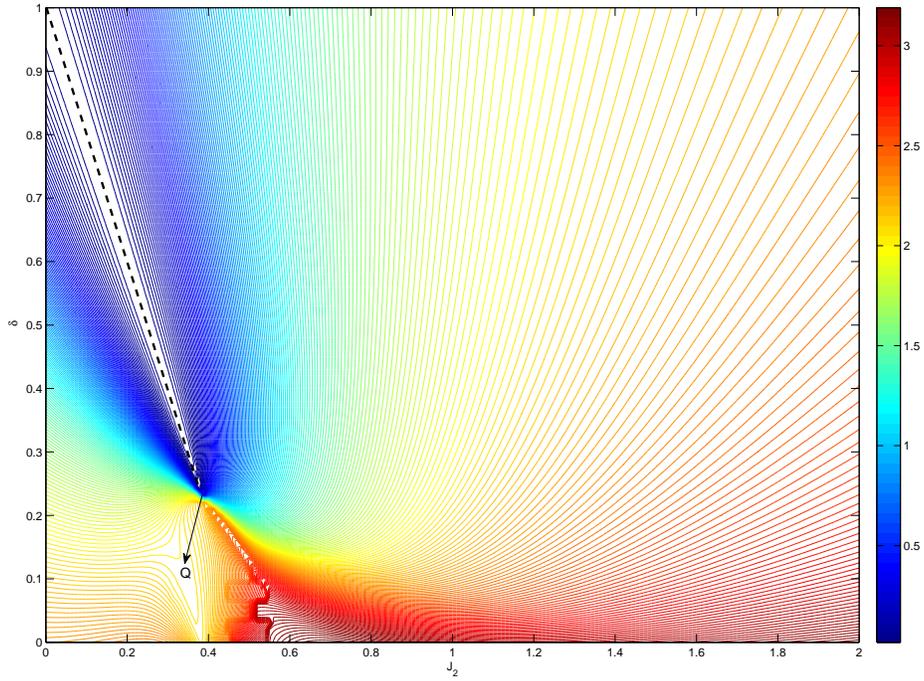}}
\caption{\small Ground state entanglement entropy contour plot of a spin-3/2
chain with 12 sites, with $J_2$ going from 0 to 2 and $\delta$ going from
0 to 1.} \label{spin3by2_contour1} \end{center} \end{figure*}

In curve (i) of Fig. \ref{spin1_derv1}, we plot $\delta_{cal}$ versus $J_2$ 
for the points corresponding to the minimum value of the entropy in the 
$J_2-\delta$ plane; this curve goes from the point `P' to the point (0, 1). 
This curve is seen to follow the line $2J_2+\delta=1$ as expected; 
we see a step staircase instead of a straight line because we have 
calculated the entropy for discrete values of $J_2$ and $\delta$. 
To detect the gapless phase along the lines `a' and `c', we calculate the 
first order derivative of the entropy (using the three-point differentiation 
formula) along the $\delta$-axis for different values of $J_2$ in this plane.
In curve (ii) of Fig. \ref{spin1_derv1}, we plot $\delta_{cal}$ versus $J_2$ 
corresponding to the maximum of the absolute value of the derivative. 
This curve follows the lines `a' and `c' closely. 

\subsubsection{Spin-1 chain fidelity and gap}

We calculate the ground state fidelity of spin-1 chains with finite sizes 
in the $J_2-\delta$ plane. For small values of $\delta$, the ground states 
of finite size systems have multiple energy level crossings with the excited 
states. Because of these finite size effects, the fidelity of the ground 
states is not a reliable tool for studying phases in regions II and III. 
However in the regions I and IV, we find no energy level crossings 
in the ground states of finite systems. We calculate the ground state 
fidelity for systems with 6, 8, 10, 12 and 16 sites for different values of 
$\delta$ along the $J_2$ axis in this plane. We find no sudden changes in 
the fidelity. We show the plot of $J_2$ versus $\delta_{cal}$ corresponding 
to the minimum fidelity for a chain of 16 sites (see inset of Fig. 
\ref{spin1_derv1}). This curve follows the lines `a', `b', `$\overline{b}$' 
and `c' qualitatively separating regions II and III from regions I and IV.

To understand the quantum phase transition on the line `b' in Fig.
\ref{phase_diag}, we calculate the spin gap of the system near this line 
(see Fig. \ref{spin1_energy_lineb}). Since the ground state of an open chain 
with spin-1 is four-fold degenerate in the regions II and III, we calculate 
the excitation energy gap as the difference between the lowest energy state in 
the $M_s=0$ sector and the first excited state in the $M_s=1$ sector. 
Our current calculations based on a finite size analysis shows that the gap 
could vanish in the thermodynamic limit. This improves our earlier report 
which had convergence difficulties \cite{pati1,pati2}. 
We also studied the behavior of the gap 
across the lines `b' and `$\overline{b}$'. For example, for the 16 site chain, 
the gap is minimum at $\delta \simeq$0.13 at $J_2=0.4$ and $\delta \simeq$0.09 
at $J_2=0.5$ (see insets of Fig. \ref{spin1_energy_lineb}). Also there is a 
change in the behavior of the gap versus 1/N as $\delta$ varies; for $\delta$ 
lying below the phase transition line, the gap saturates to a finite value, 
while for $\delta$ lying above the line, the gap continues to decrease 
steadily as $N$ increases.

The line `b' appears to be a phase transition line which separates the Haldane 
and spin Peierls phases which are both gapped. These two phases differ in 
several ways. For an open chain, the ground state has a four-fold degeneracy 
(a spin singlet and a spin triplet which are degenerate) with spin-1/2 states 
at the ends in the Haldane phase, but is non-degenerate (spin singlet) in the 
spin Peierls phase. Further, the Haldane phase has a non-local string 
order parameter \cite{nijs}.

\subsection{The spin-3/2 system in $J_2-\delta$ plane}

In this section we study the phase diagram of the spin-3/2 system in the 
$J_2-\delta$ plane. To the best of our knowledge, the quantum phases of 
the Heisenberg spin-3/2 antiferromagnetic chain in the $J_2-\delta$ plane 
has not been studied earlier. But from the field theory analysis of the spin 
chain, half-odd integer systems are gapless at $\delta=0$ and for small values
of $J_2$. With dimerization ($\delta \ne0$), it is predicted that spin-3/2 
system should be gapless at $\delta=2/3$ for $J_2=0$ \cite{pati2,affleck}. 

\subsubsection{Spin-3/2 entropy phase diagram (contour plot)}

\begin{figure}[]
\begin{center} {\includegraphics[width=8.8cm]{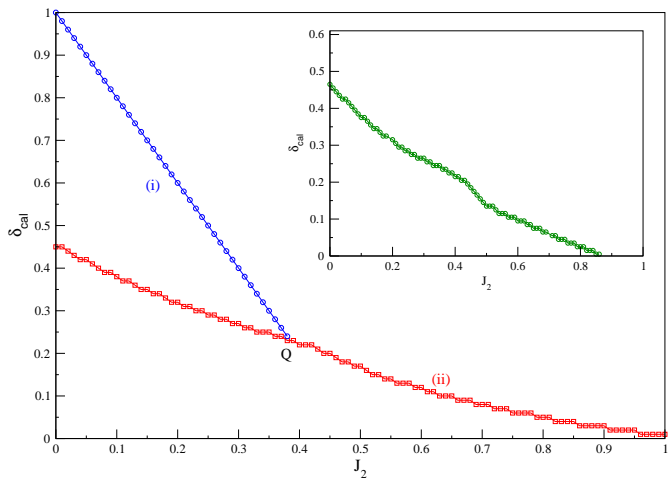}}
\caption{\small Curve (i) shows the $(J_2,\delta_{cal})$ values (squares) 
corresponding to the minimum entropy. Curve (ii) shows the 
$(J_2,\delta_{cal})$ values corresponding to the maximum absolute values 
of the first order derivative of the entropy w.r.t. $\delta$. 
Inset shows the points corresponding to minimum fidelity (taking $\delta$ 
as the variable parameter). All the results are obtained for the 12 site 
spin-3/2 chain.} \label{spin3by2_derv1} \end{center} \end{figure}

\begin{figure*}[]
\begin{center} 
{\includegraphics[width=14.0cm]{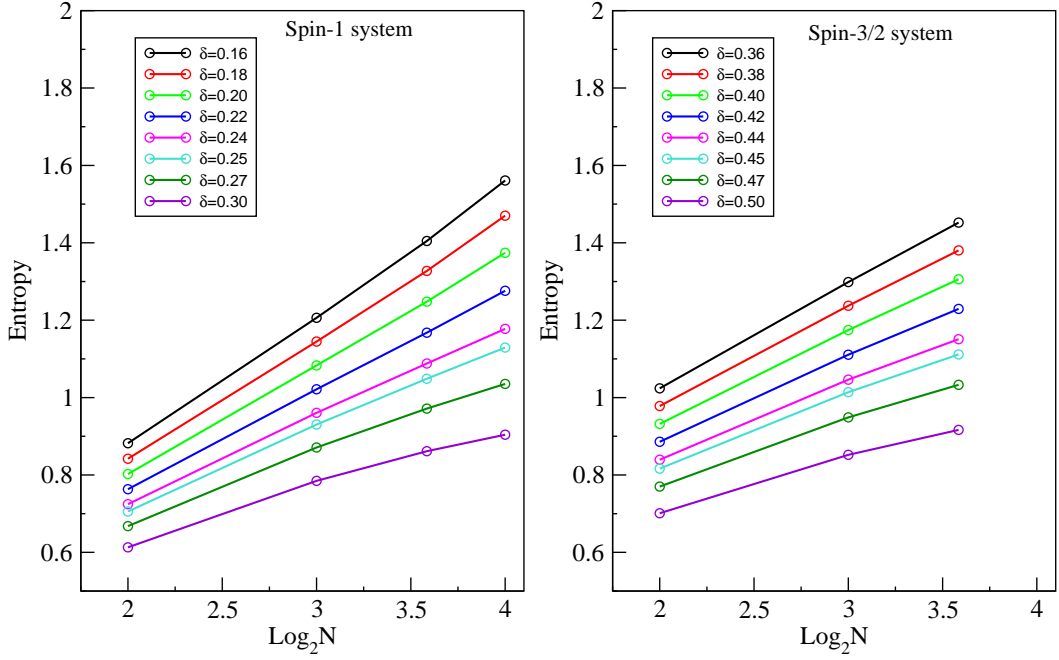}}
\caption{\small For spin-1 and 3/2 systems, the entropy versus 
logarithm of the system size is shown for different $\delta$ values with
$J_2 = 0$.}
\label{scaling1_spin1_3by2}  
\end{center} \end{figure*}

We study different quantum phases of spin-3/2 chain. We use spin parity 
symmetry to break the degeneracy (within numerical accuracy) of the ground 
state of this system. For 12 site chain, a contour plot of the entropy 
is shown in Fig. \ref{spin3by2_contour1}. As in the spin-1/2 and spin-1 
cases, we observe an order-disorder transition along $2J_2+\delta=1$ line. 
The line starts approximately from the point Q = ($J_2=0.38$, $\delta=0.24$) 
and extends up to ($J_2=0$, $\delta=1$). For small $J_2$ values near 
$\delta$ = 0 in the contour plot, 
the line pattern is similar to that for spin-1/2 case and quite different 
from the spin-1 case. This suggests that there can be a gapless phase at 
$\delta=0$ as predicted. For $J_2$ = 0, the line density is very high between 
$\delta=0.4$ and 0.5; this suggests another gapless phase in this region as 
predicted by field theory. As in the spin-1 case, the density of lines is 
high at larger $J_2$ values (about $J_2=1$). This suggests a numerically 
gapless phase in this region. We also confirm these gapless phases by 
comparing numerical energy gaps in those regions with that of a phase at 
large values of $J_2$ and $\delta$ where the density of lines is very low.
For better understanding of these quantum phases 
we plot $\delta_{cal}$ versus $J_2$ values corresponding to 
the minimum entropy (curve (i) from Fig. \ref{spin3by2_derv1}) 
above the `Q' point. This curve follows the $2J_2+\delta=1$ line similar 
to the spin-1/2 and spin-1 cases. We also plot the points corresponding to 
the maximum absolute values of the first order derivative of entropy with 
respect to $\delta$; this is shown as curve (ii) in Fig. \ref{spin3by2_derv1}.
This is similar to the curve representing a numerically gapless phase for 
a spin-1 system (curve (ii) in Fig. \ref{spin1_derv1}). This suggests that 
there can also be a gapless region along the curve (ii) for spin-3/2 system. 
The gapless point ( curve (ii) of Fig. \ref{spin3by2_derv1}) at $\delta=0.45$ 
at $J_2=0$ is consistent with the value $\delta=0.431$ reported 
\cite{nakamura}. Note that this value is different from the field theory 
prediction of 2/3 \cite{pati2,affleck}.

We calculate the ground state fidelity of the 12 site spin-3/2 chain along 
different $J_2$ values with $\delta$ as the variable parameter. We plot $J_2$ 
versus $\delta$ correspond to minimum fidelity in this plane (see inset of 
Fig. \ref{spin3by2_derv1}). The curve approximately follows the curve (ii) 
in Fig. \ref{spin3by2_derv1}. 

To further investigate the gapless points which are expected to occur 
at certain non-zero values of $\delta$ at $J_2=0$ for the spin-1 and 
3/2 systems, we have shown the entropy versus the logarithm of the system 
size for different $\delta$ values in Fig. \ref{scaling1_spin1_3by2}. 
The present system size appears to be too small to numerically verify 
the conformal field theory prediction \cite{calabrese09} of 
$S=\frac{c}{6}\log_2 N$ (with $c=1$) at the critical points which occur
at certain values of $\delta$ (numerically estimated to be
$0.24$ for spin-1 and $0.43$ for spin-3/2).

\section{Conclusion}

We have used entanglement entropy and fidelity as tools to study the 
different quantum phases and quantum critical regions of the spin-1/2, 
1 and 3/2 chains in the $J_2-\delta$ plane. 
For this study, we have employed extensive exact diagonalization of spin 
chains with up to 20 sites depending on the site spin. We have considered 
201 values of $J_2$ in the range 0 to 2 and 101 values of $\delta$ in the 
range 0 to 1 corresponding to over 20,000 grid points.

We have studied the complete phase diagrams of these three systems using 
entropy contour plots and fidelity in the $J_2-\delta$ plane. 
We have been able to identify the quantum phase transitions 
from gapless to gapped phases using the density of the contour lines of 
the entropy and the minimum fidelity. Though the full phase diagram of the 
spin-3/2 system has not been investigated before, we have conjectured the 
existence of some gapless regions and an order-disorder line by studying its 
phase diagram and comparing it with the phase diagrams of the spin-1/2 and 
spin-1 systems. Our main results are that we find indications of a gapless 
region near $\delta=0$ and small values of $J_2$ in the spin-1/2 system, 
a gapless region at finite $\delta$ in the spin-1 system (lines `a' and `b' 
in Fig. \ref{phase_diag} (b)), and two gapless regions near $\delta=0$ and 
around $\delta \sim 0.4 ~-~ 0.5$ for $J_2=0$ in the spin-3/2 system.

\section{Acknowledgment}

S. R. and D. S. are thankful to the Department of Science and Technology 
(DST), India for financial support through various projects.

\end{document}